# Evidence of Pure Spin Current

Weiwei Lin and C. L. Chien

*Department of Physics and Astronomy, Johns Hopkins University, Baltimore, Maryland 21218, USA*

*Email: wlin@jhu.edu; clchien@jhu.edu*

Evidences of pure spin current are indistinguishable from those of many parasitic effects. Proper choices of materials and methods are essential for exploring pure spin current phenomena and devices.

A pure spin current has the unique attribute of delivering spin angular momentum with a minimal of charge carriers, thus generating least amount of Joule heat as an electronic spin current in a metal and negligible heat as a magnonic spin current in an insulator. Pure spin current phenomena were first explored theoretically and experimentally in the 1970's by Dyakonov, Perel, Monod, Janossy, Silsbee, and others, with the realization of spin Hall effect (SHE)[1], spin pumping (SP)[2,3], non-local lateral spin transport[4], and thermal spin injection in rapid succession. In recent years, there has been intense interest in exploring pure spin current phenomena and switching of magnetic devices using spin orbit torques. The main schemes for generating pure spin current remain the same, but the exploration has been extended beyond metals and ferromagnets, to materials that include antiferromagnets[5], topological insulators, two-dimensional materials, and materials that harbor Rashba-Edelstein effect[6]. However, the large variances of key parameters, including some claims of spin Hall angle larger than 1[6], suggest that the evidence of pure spin current requires closer examination.

In SHE, a charge current $j_C$ in a non-magnetic metal with strong spin-orbit coupling (SOC) generates a pure spin current $\mathbf{j}_S \propto \theta_{SH} \mathbf{j}_C \times \boldsymbol{\sigma}$ in the perpendicular direction, with the spin index $\boldsymbol{\sigma}$ in the transverse direction perpendicular to both $j_C$ and $j_S$, where $\theta_{SH}$ is the spin Hall angle with a magnitude less than 1, specifies the efficiency of the charge-spin conversion. Some heavy metals (HMs) have large positive (Pt and Pd) or negative (W and Ta) $\theta_{SH}$ values[7,8], but all must adhere to $|\theta_{SH}| \leq 1$. More often, one uses a ferromagnetic (FM) material to inject a pure spin current by SP via ferromagnetic resonance (FMR)[7,8], or by longitudinal spin Seebeck effect (LSSE) via a temperature gradient[5]. In both cases, the direction of the FM magnetization **M** sets the spin index $\sigma$ of the pure spin current.

Since a pure spin current defies detection by the usual electrical means, one indirectly measures the pure spin current after converting it into a charge current or a DC voltage by the inverse spin Hall effect (ISHE)[7,8] using a HM with a substantial $\theta_{SH}$. Although the ISHE process would give rise to a voltage in the detecting HM layer, a voltage measured in an actual experiment may *not* be exclusively that of spin-to-charge conversion. Ascertaining pure spin current contribution is one of the essential challenges in pure spin current phenomena. Consider the numerous SP experiments in Pt/Py, the most widely studied system, where permalloy (Py) = $Ni_{81}Fe_{19}$ is a FM metal, the reported values of the ISHE voltage and $\theta_{SH}$ span more than two orders of magnitude[9,10]; sometimes with the opposite sign for the same metal, e.g., in Ta/Py[10] and Ta/YIG[8]. In SP experiments in NM/FM metal bilayers, there are many parasitic effects[11,12]. The microwave driven GHz magnetization precession at the FMR induces not only rf charge current but also a dc current due to rectification[11,12], leading to a spin-polarized current rather than a pure spin current. As a result, anisotropic magnetoresistance (AMR), anomalous Hall effect (AHE), and planer Hall effect, etc., also contribute to the voltage, but their contributions in actual SP experiments are difficult to extract[11,12].

In addition to the rectification effects, unintentional but unavoidable heat generation in FMR has been noted since the 1970's[13] but rarely taken into account. Recently, Yamanoi *et al.*[14] show that FMR generates a very large temperature difference across the sample of as much as $\Delta T = 12$ K, which is larger than those in most thermal spin injection experiments. Furthermore, the temperature gradient increases with the microwave power, thus unavoidable in any SP experiment[14]. As a result, all the heat-related contributions, not only anomalous Nernst effect (ANE) in FM metals[15] and LSSE in FM insulators[5], but also Nernst effect in the cases of semimetals and semiconductors, must be carefully analyzed, further acerbating the already complex situation in ISHE detection. These parasitic effects, would greatly complicate if not preclude, reliable extraction of the parameters related to the spin current in the SP experiments. Some attempts, e.g., angular and thickness dependences, have been made to extract the spin-to-charge voltage from those of the extrinsic microwave effects[12]. However, certain prevailing parasitic effects have the same angular dependence as that of ISHE, thus inseparable.

We show in this work that the parasitic contributions, not only substantial, can even overwhelm that of the pure spin current. We propose and demonstrate several criteria with which one can validate the spin current contributions. More importantly, we show that the FM metals are unreliable, whereas the FM insulators are far better, pure spin current injectors. Finally, we address the key question of whether coherent magnetization precession (resonance) injects pure spin current in the SP experiments as theoretically proposed.

In most pure spin current phenomena and devices, one employs HM/FM bilayers deposited on a substrate. A pure spin current $\mathbf{j}_S$, driven across the HM/FM interface, is detected by the ISHE in the HM layer. As such, the sign of the ISHE voltage must adhere to that of $\theta_{SH}$ of the HM. If the pure spin current has been blocked, or if there is no HM, there should be no ISHE voltage. Thus, the minimal criteria we propose include; (1) Two HMs with opposite $\theta_{SH}$ must exhibit ISHE voltages of opposite signs. (2) The ISHE voltages must vanish when an insulating layer (e.g., MgO, SiO$_x$) has been inserted in the structure to block $\mathbf{j}_S$. (3) There should be no ISHE voltage without the presence of the HM layer. This seemingly trivial criterion is designed to reveal parasitic effects unrelated to ISHE. Also essential, when the "pure spin current" has been generated by motion of charge carriers, including both SP and LSSE experiments, heating is unavoidable and its consequences must be addressed. Because HM/FM bilayer has been deposited on much thicker substrate, any heat generation would cause an out-of-plane temperature gradient $\nabla_z T$ [15], whose consequence must be determined.

We first examine the pure spin current injection in HM/Py bilayers, the systems of numerous SP experiments. We use thermal spin injection by applying a temperature gradient $\nabla_z T$, in the out-of-plane direction to highlight the heat related parasitic effects. The ISHE, if present, would generate an electric field of $\mathbf{E}_{ISHE} \propto \mathbf{j}_S \times \boldsymbol{\sigma}$, where $\mathbf{j}_S$ is along $\nabla_z T$ and $\boldsymbol{\sigma}$ is along the direction of the in-plane magnetization (**M**) of Py as shown in Fig. 1a. As shown in Fig. 2a, under a temperature difference of $\Delta T = 10$ K near room temperature, one observes a voltage in Pt/Py, ($\theta_{SH} > 0$ for Pt) that reaches a saturated value when **M** of Py is aligned, and the voltage changes sign when **M** reverses. The coercive field of the Py layer defines the loop width. These are the expected signatures of spin-to-charge conversion via ISHE. However, if this is so, a HM with $\theta_{SH} < 0$, such as W, must show a voltage of the opposite sign. However, as shown in Fig. 2a, a voltage of the *same* sign has been observed in Pt/Py and W/Py, even though Pt and W have $\theta_{SH}$ of *opposite* signs, vividly demonstrating the violation of criterion (1), thus *neither* result is due to spin-to-charge conversion. In fact, this is the ANE occurring in FM Py that generates an electric field of $\mathbf{E}_{ANE} \propto \mathbf{j}_C \times \boldsymbol{\sigma}$, where $\mathbf{j}_C$ is also along $\nabla_z T$ as shown in Fig.

1b. These analyses show that the ISHE and ANE voltages share the same symmetry and angular dependence, hence inseparable and additive. In fact, a voltage of the same characteristics has already been observed in Py alone, due entirely to ANE, in violation of criterion (3). An additional HM layer of either Pt or W only reduces the ANE voltage through shunting. These experiments show clearly that in HM/Py, the ANE in Py completely dominates the measured voltage. There is *no* measurable spin-to-charge conversion in HM/Py. Similar results have also been observed HM/FM using other FM metals of Fe and CoFeB[16].

SP experiments have also been performed on Bi[6] and Bi materials, such as $Bi_2Se_3$, which are some of the best thermoelectric materials with additional complications with large heat-generated electrical responses. Due to the presence of semimetal/metal interface, the rectification effects in the Bi/Py bilayer are more pronounced. In addition to the rectification effects and the ANE in Py, there is also the significant Nernst effect in the Bi materials. The Nernst effect generates an electric field $\mathbf{E}_{NE} \propto \mathbf{j}_C \times \mathbf{H}$, where $\mathbf{j}_C$ ∥$\nabla_z T$ and $\mu_o \mathbf{H}$ is the in-plane magnetic field. Comparing with $\mathbf{E}_{ISHE} \propto \mathbf{j}_S \times \boldsymbol{\sigma}$, the Nernst voltage and ISHE voltage have the same angular dependence and hence additive. The Nernst effect is negligible in common metals, but $10^6$ to $10^9$ times larger in Bi materials due to the low carrier density, high mobility, and small Fermi energy. Since SP requires an external magnetic field to align the in-plane magnetization, a substantial Nernst voltage would result but that is unrelated to ISHE. If the measured voltage in the SP experiments were attributed only to ISHE, such as in Bi/Ag/Py, unphysical values of $\theta_{SH}$ > 1 would have been concluded[6].

These experiments clearly show that FM metals are not reliable pure spin current injectors. The culprits are the abundant charge carriers that respond to all stimulations, electromagnetic (rectification, AMR, AHE, etc) and thermal (ANE, Nernst effect, etc), resulting in various voltages, some of which are inseparable from, and indeed overwhelm, that of the ISHE[11,12]. Since most of the parasitic effects are the results of the movement of charge carriers, one can rid these effects by eliminating the charge carriers. As we show, the FM insulators, such as $Y_3Fe_5O_{12}$ = YIG, with no charge carriers are far better pure spin current injectors[5,7,8]. In addition, YIG is also well known to have the lowest damping of magnetic materials, metals or insulators[7,8]. In the following, we discuss spin-to-charge conversion using YIG.

In Fig. 2b, we show the results of thermal spin injection in Pt/YIG and W/YIG, which exhibit voltages of opposite signs, thus satisfying criterion (1). When a 5 nm thick MgO layer has been inserted at the HM/YIG interface to block the passage of the spin current, the voltage indeed vanishes, thus fulfilling criterion (2). Since YIG is electrically insulating, there is no spurious voltage without the HM layer. The results of Pt/YIG and W/YIG show a switching loop at low fields. This is due to the domain structure of bulk YIG substrates unrelated to spin current[17]. Indeed, YIG has been used to inject pure spin current into Au, Ta, Cr, Py, Au-Ta alloys, etc., with consistent results and always revealing the correct sign of the $\theta_{SH}$ of the detected metal. After taken into account spin diffusion length and spin mixing conductance, one can quantitatively determine $\theta_{SH}$. We show that pure spin current phenomena can be reliably pursued using YIG as a spin current injector in thermal injection. There have also been SP reports using HM/YIG[7,8].

With the absence of most parasitic effects, the results are much more consistent, including revealing the correct sign of $\theta_{SH}$ of the HM.

We next address the origin of spin pumping. When SP was first explored in the 1970's, the Monod, Janossy and Silsbee (MJS) model considers microwave driven electron spin diffusion[2,3]. In 2002, Tserkovnyak, Brataas, and Bauer (TBB) proposed coherent SP by FMR that occurs at a fixed frequency, involving GHz *coherent* precessing of magnetization of a ferromagnet transferring spins (pure spin current) into an adjacent normal metal layer[18]. This leads to an enhanced Gilbert damping (torque) of the FM layer due to spin backflow from the interface[18]. Coherent SP is also simpler to model thus more appealing. While some SP experiments have subscribed to coherent SP, others have argued that the observed enhancement of Gilbert damping in NM/FM bilayers may have resulted from microwave induced heating effects[19] and the modified magnetization and magnetic anisotropy of the FM layer and NM due to magnetic proximity effects[20]. Recently, it has been found that there is no relationship between the spin mixing conductance deduced from the enhancement of Gilbert damping and the ISHE voltage[21]. These results indicate that the enhanced damping and the measured ISHE voltages in the SP experiment may not be attributable to spin current.

We seek experimental evidence for coherent magnetization precession (resonance) induced pure spin current. One unique characteristic of coherent spin current is that it is temperature independent[18]. This feature of coherent spin current is in sharp contrast to that of incoherent thermal magnon current, which reduces with decreasing temperature, and vanishes at zero temperature[5,22]. Thus, the telltale sign of coherent spin injection can be unequivocally confirmed by SP experiment conducted near $T = 0$ K. Such experiments in Pt/YIG have recently been available.

We first describe the results of incoherent thermal spin injection via LSSE that we have made in Pt/YIG. At low temperatures, the measured ISHE voltage decreases with decreasing temperatures and, by extrapolation, vanishes at $T = 0$ K as shown in Fig. 3a. This is the expected behavior of incoherent thermal magnons corresponding to a wide spectrum of frequencies[5]. Others have also observed similar results with vanishing contribution near $T = 0$ K[22]. There are also recent reports of SP in Pt/YIG over a wide temperature range with the results[23] adapted in Fig. 3b. Most importantly, the ISHE voltage from SP in Pt/YIG steadily decreases with decreasing temperatures and extrapolated *also* to zero at $T = 0$ K. Since the resistivity of the thin HM layer has very weak temperature dependence and never vanishes, the strong voltage dependence is essentially that of the injected spin current after spin-to-charge conversion. Vanishing pure spin current at $T = 0$ K is decidedly not the predicted behavior of coherent spin injection driven by FMR. Previously, FMR in YIG at $T \approx 0$ K has been demonstrated[24] and that the spin Hall angle and resistivity of Pt remain finite. Thus the null SP results at $T \approx 0$ K in Fig. 3b shows the absence of experimental evidence for the generation of coherent pure spin current via resonant magnetization precession in YIG. It remains an experimental challenge to capture the elusive theoretically predicted coherent spin injection by FMR.

In summary, because the evidence of pure spin current injection can only be indirectly measured by spin-to-charge conversion via the ISHE, the evidences need to be carefully analyzed. We propose several criteria with which one can experimentally scrutinize and validate the results. We show that FM metals, such as Py, despite being used in many SP experiments, are plagued with a barrage of parasitic effects, hence unsuitable as pure spin current injectors. The overwhelming ANE in FM metals renders the contribution from pure spin current undetectable. Since most of the parasitic effects originate from charge carriers in FM metals, FM insulators, such as YIG, are far better injectors for pursuing pure spin current phenomena as we have experimentally demonstrated. Finally, to date, only incoherent spin injection has been observed, with no experimental evidence of coherent SP.

**Acknowledgements**

This work was supported by National Science Foundation DMREF, Grant No. 1729555. W.L. was supported in part by SHINES, Grant No. DE-SC0012670, an Energy Frontier Research Center and by Grant No. DE-SC0009390, both from the U.S. Department of Energy, Office of Science, Basic Energy Science.

**Figure Legends**

Figure 1  (a) Schematic of longitudinal spin Seebeck effect (LSSE) under a vertical temperature gradient $\nabla_z T$ in a normal metal(NM)/ferromagnetic insulator (FMI) structure generating an electric field of $\mathbf{E}_{\text{ISHE}} \propto \mathbf{j}_S \times \boldsymbol{\sigma}$, where the spin current $j_S$ is along $\nabla_z T$ and $\boldsymbol{\sigma}$ is along the direction of the in-plane magnetization (**M**). (b) Schematic of anomalous Nernst effect (ANE) in a ferromagnetic metal (FM) layer on an insulating substrate generating an electric field of $\mathbf{E}_{\text{ANE}} \propto \mathbf{j}_C \times \boldsymbol{\sigma}$, where the charge current $\mathbf{j}_C$ is along $\nabla_z T$ and $\sigma$ is along the direction of the in-plane **M**. (c) Both the LSSE and the ANE give voltages with the same symmetry, thus indistinguishable and inseparable. The result is the ANE in a 15 nm thick Py film at room temperature under a vertical temperature difference $\Delta T$ of 10 K.

Figure 2  (a) Magnetic field dependence of voltages in Py(15 nm), Pt(2.5 nm)/Py(15 nm), and W(2.5 nm)/Py(15 nm) films at room temperature under a vertical temperature difference of $\Delta T$ =10 K applied between the film surface and the bottom of the 0.5 mm thick SiO$_x$/Si substrate. (b) Magnetic field dependence of voltage in the Pt(3 nm)/YIG, W(3 nm)/YIG and Pt(3 nm)/MgO(2 nm)/YIG samples at room temperature under $\Delta T$ =10 K applied between the film surface and the bottom of the 0.5 mm thick polycrystalline YIG slab.

Figure 3  (a) Temperature dependence of the transverse thermopower $S = \frac{V_{\text{ISHE}}/L}{\Delta T/t_{\text{YIG}}}$ in a Pt(3 nm)/polycrystalline YIG(0.5 mm) slab under fixed length (*L*) of the Pt layer, thickness of YIG ($t_{\text{YIG}}$), and temperature difference ($\Delta T$) across the sample. *S* is proportional to $V_{ISHE}$. (b) Temperature dependence of the voltage/power ratio in a Pt(6 nm)/YIG(10 µm) bilayer on a $Gd_3Ga_5O_{12}$ substrate measured using spin pumping, adapted from ref. 23, Macmillan Publishers Ltd. The line is a guide to the eye that extrapolates to zero at zero temperature.

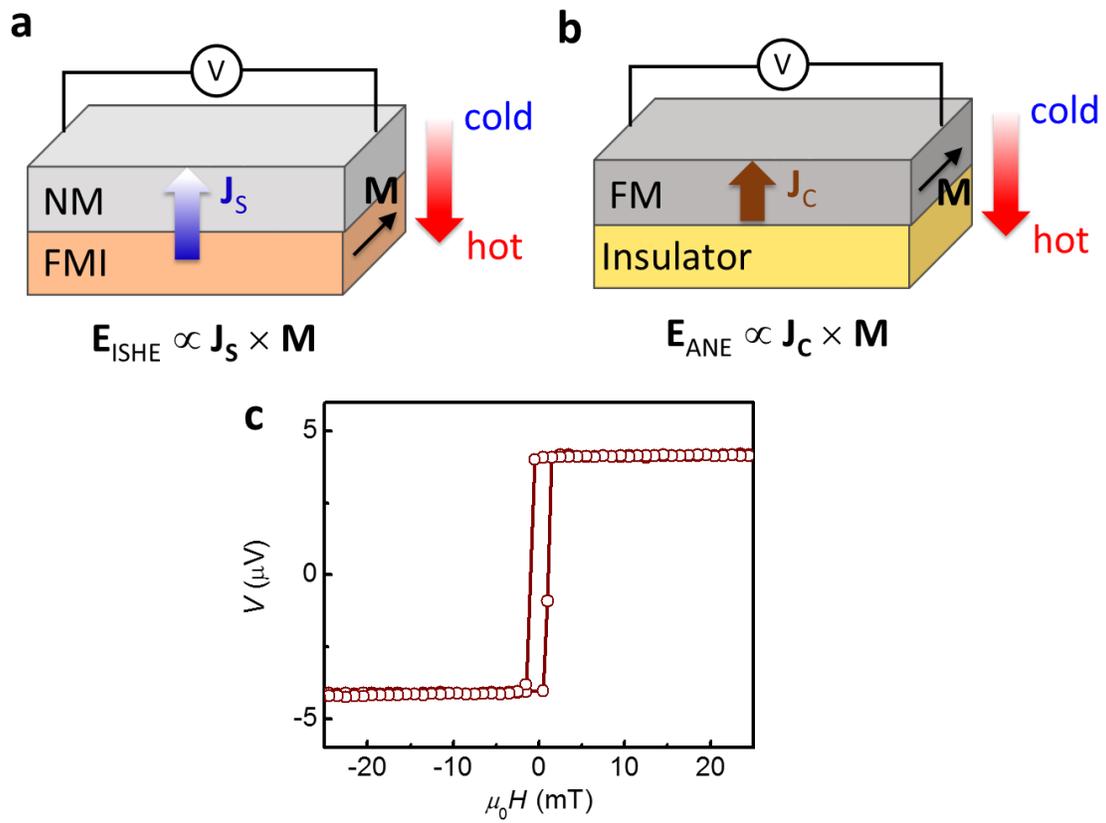

Figure 1

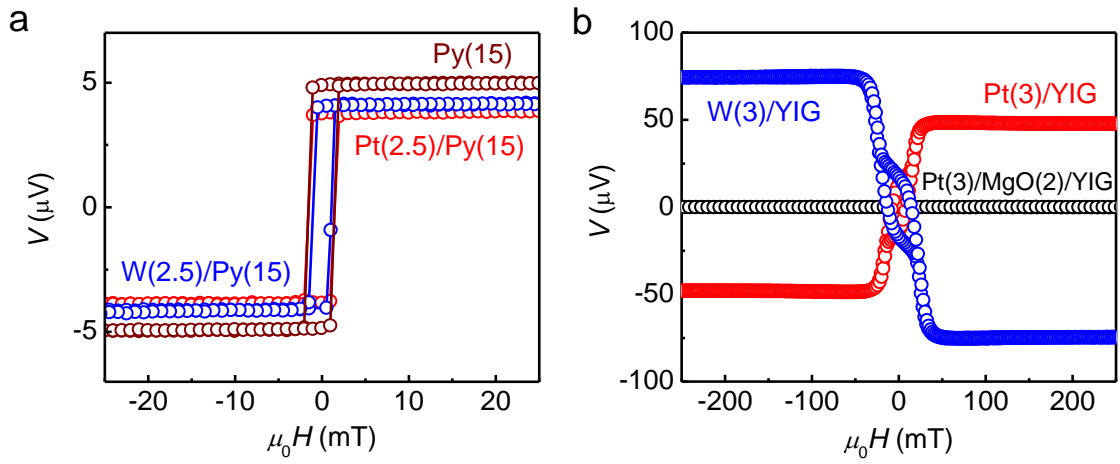

Figure 2

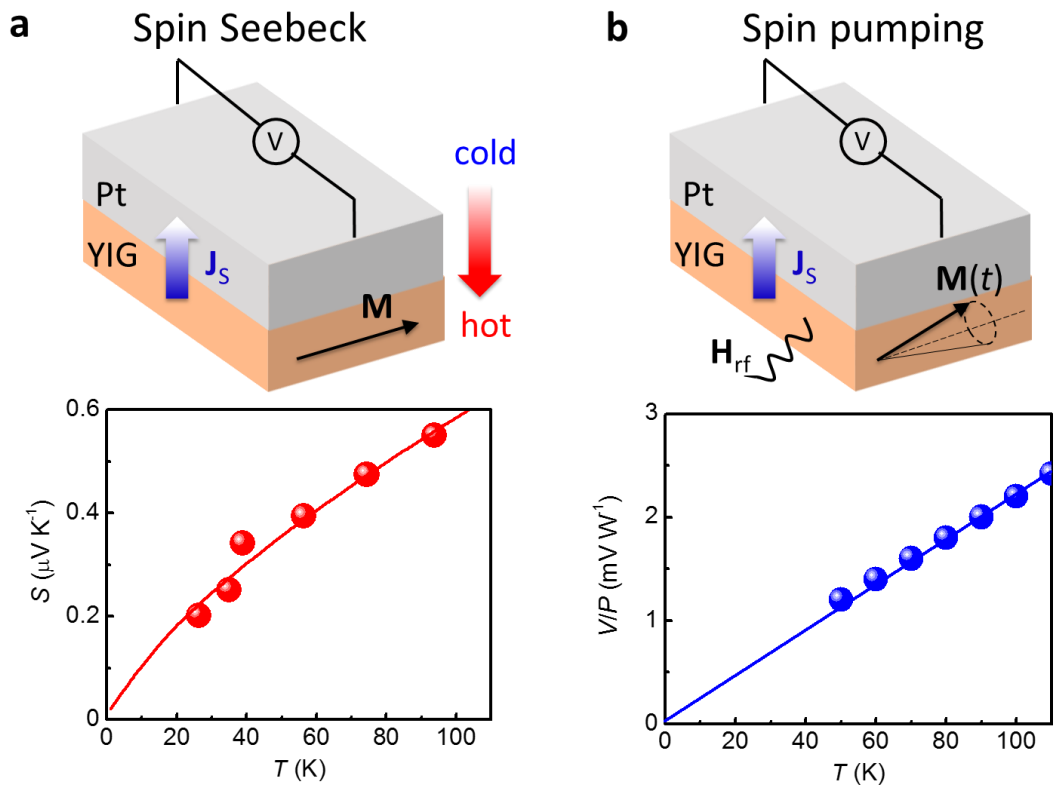

Figure 3